# Plasmon-induced photonic switching in a metamaterial

H. Xu and B. S. Ham

Center for Photon Information Processing, and the Graduate School of Information and Communications,
Inha University, Incheon 402-751, Republic of Korea

Using light-induced localized surface plasmon interactions in a metamaterial, we present a plasmonic control of light absorption for photonic switching. We discuss that the present surface plasmon-induced photonic switching is comparable with coherence swapping in a tripod optical system based on electromagnetically induced transparency. This outcome opens a door to active controls of the surface plasmons in a metamaterial for potential applications of nano photonics.

Surface plasmon polaritons (SPPs) have been intensively studied for the last several years. From a simple transmission scheme of SPPs in a narrow metal stripe interfacing with a dielectric material [1] to complicated device applications of the SPPs [2], metamaterial has been intensively studied for a wide range of frequencies, from terahertz to optical regime. Researchers have also paid attention to localized surface plasmons for interesting physics of resonance and fluorescence [3,4]. Plasmonic mode control through use of light is an emerging research area toward light-controlled nano optics in metamaterials. A recent observation of optical field-based SPP switching is a good example of light controlled plasmonics, where the physics lies in saturation phenomenon [5]. The saturation phenomenon, however, is limited in the strong field. Nano photonics inherently should work at the low power limit.

Electromagnetically induced transparency (EIT) is a direct result of destructive quantum interference between two pathways induced by another light [6-8]. Unlike most photonic devices, EIT uses quantum interference, where zero absorption probability at line center is induced from the destructive quantum interference [9]. Thus, EIT free from the saturation phenomenon has been applied for ultraefficient nonlinear quantum optics [10,11]. Group velocity control by use of EIT has also been intensively studied for fundamental physics [12, 13] as well as various applications such as entanglement generation [14] and photon logic gates [15]. Clearly, EIT greatly benefits surface plasmon-based nano photonics that requires low light power.

Recently an EIT-like feature has been suggested for the plasmonic mode control using localized surface plasmons in a metamaterial model [16]. In this Letter we propose a plasmonic mode control for optical switching in a metamaterial, where the plasmon-induced optical switching resembles quantum coherence swapping in an EIT-based tripod system [17]. Here we intuitively modify the plasmon-induced EIT-like model of Ref. 16. The phenomenon of EIT-based switching has been demonstrated in both N-type [18] and tripod optical systems [17,19], where the system requires a third control light in addition to the EIT requirements. In a tripod system, the EIT-induced single dark resonance is swapped for another dark resonance when a third light is applied. This effect is called coherence swapping [17,19] or simply double dark resonance [18,20]. As a result, at line center of EIT, the quantum interference-based absorption cancellation results in a strong absorption [17-21]. Even though the present scheme is more passive, it gives a fundamental basis for an active plasmonic control in a scalable plasmon-based nano device using multi-channel optical frequencies.

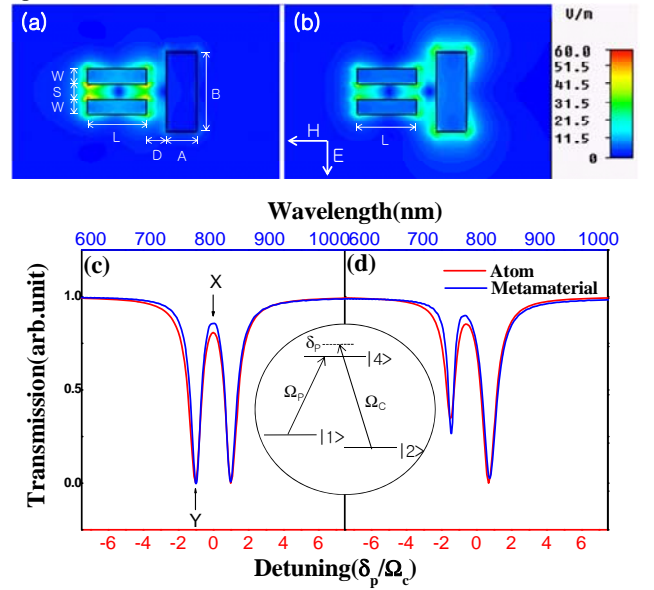

FIG. 1 (color online) Numerical simulations of localized surface plasmon distribution on a metamaterial interacting with vertically incident light for (a) dark mode ($\lambda$=796.5 nm; indicated by "X" in (c)) and for (b) bright mode ($\lambda$=766 nm; indicated by "Y" in (c)). Numerical simulations for the dark mode with (c) no detuning ($L_0$=L, $\delta_C$=0) and (d) with detuning ($L_0$=L−8nm, $\delta_C$=3/8$\Omega_C$): Blue (red) curve represents for the metameterial (atomic model). For the atomic model, a $\Lambda$−type system interacting with resonant light $\Omega_C$ and $\Omega_P$ is considered. A=60 nm; B=160 nm; L=118 nm; D=40 nm; W=30 nm; S=30 nm; The thickness of each strip is 20 nm; $\Omega_P$=1/20$\Omega_C$; $\gamma_{31}$= $\gamma_{32}$=1/4$\Omega_C$; $\gamma_{12}$=1/8$\Omega_C$.

Figures 1(a) and 1(b), respectively, show dark and bright modes of the localized plasmon interactions induced by a monochromatic light vertically incident on the metamaterial. The metamaterial scheme corresponds to EIT in a Λ−type system interacting with two resonant optical fields as proposed in Ref. 16. Here the dark (bright) mode stands for nonabsorption (absorption) resonance in EIT terminology, where the nonabsorption (absorption) resonance is due to destructive (constructive) quantum interference between two dressed states induced by the coupling light. The inset of Fig. 1(c) represents a corresponding atomic level structure, and Fig. 1(c) shows each simulation: The red curve represents the atomic model with no coupling detuning ($\delta_C=0$), and the blue curve represents the metamaterial at far field detection. The parallel double metal stripes represent a coupling field ($\Omega_C$), while the vertical single metal stripe represents a probe field ($\Omega_P$). The length and width of the metal stripes affect transition frequency and linewidth, respectively (More detailed discussions will be presented elsewhere). For example, a red (blue) shift occurs in the transition frequency if the length of the metal stripe increases (decreases). In the plasmon interaction model of Fig. 1, the length modification of the parallel metal stripes determines the coupling field detuning $\delta_C$ as shown in Fig. 1(d). The polarization of the incident light is parallel to the vertical single metal stripe as indicated.

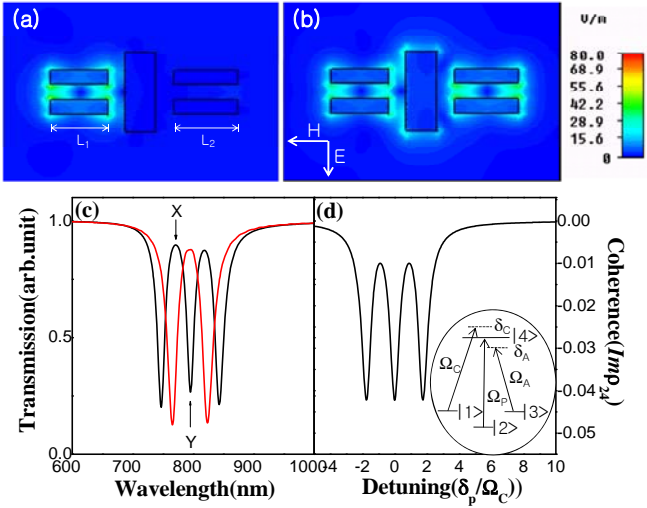

FIG. 2 (color online). Plasmon induced photonic switching based on coherence swapping. Numerical simulations of localized surface plasmon distribution on a metamaterial interacting with vertically incident light for (a) single dark mode ($\lambda$=771.4 nm; indicated by "X" in (c)) and for (b) double dark mode ($\lambda$=796.5 nm; indicated by "Y" in (c)). (c) Far field detection simulation, where the red curve is for Fig. 1(a) as a reference. (d) Numerical simulation for a corresponding atomic tripod model. All parameters are the same as Fig. 1 unless otherwise indicated: $L_1 = L - \Delta L_1$; $L_2 = L + \Delta L_2$; $\Delta L_1 = \Delta L_2 = 6$ nm; $\Omega_A = \Omega_C$; $\delta_A = -\delta_C$.

In Fig. 2, parallel metal stripes are added to the right wing of Fig. 1. The length of the double parallel stripes $L_1$ and $L_2$ changes by $\Delta L$ in an opposite direction: $L_1 = L - \Delta L$; $L_2 = L + \Delta L$. Here reduced length in L represents a blue shift in the detuning. Figures 2(a) and 2(b), respectively, show a single-dark mode and a double-dark mode of the localized plasmon interactions induced by a monochromatic light vertically incident on the metamaterial, where the scheme corresponds to a detuned tripod optical system interacting with three lights, as shown in Fig. 2(d). The double-dark mode in Fig. 2(b) results in absorption enhancement at line center. The single dark modes appear at the side bands of the double-dark mode (probe resonance transition), where each position (energy splitting) is determined by each detuning $\delta_C$ or $\delta_A$. As an N-type system [18] or a tripod system [17,19] originates at the Λ−type system of EIT [7], Fig. 2 shows a modified model of Fig. 1 for the double-dark resonance or coherence swapping. The red curve in Fig. 2(c), as in Fig. 1(c), serves as a reference for EIT, while the black curve represents a far-field detection of transmission in Fig. 2(a). In Fig. 2(c) there is an absorption enhancement at line center, where the added double parallel metal stripes ($L_2$) play a major role. Figure 2(d) represents a numerical simulation of the tripod atomic model (see the inset), which match quite well Fig. 2(c).

For the numerical calculations of the tripod model, time-dependent density matrix equations are derived from the interaction Hamiltonian using the Liouville equation [17]:

$$H = \hbar\left\{-\delta_C|1\rangle\langle 1| - \delta_P|2\rangle\langle 2| - \delta_A|3\rangle\langle 3| - \frac{1}{2}(\Omega_C|1\rangle\langle 4| + \Omega_P|2\rangle\langle 4| + \Omega_A|3\rangle\langle 4|) + H.c.\right\},$$

$$\dot{\rho} = -\frac{i}{\hbar}[H, \rho] + decay,$$

$$\dot{\rho}_{24} = -i\Omega_P(\rho_{22} - \rho_{44}) - i\Omega_A\rho_{23} - i\Omega_C\rho_{21} - i\delta_P\rho_{24} - \gamma_{24}\rho_{24},$$
(1)

where $\delta_C = \omega_{14} - \omega_C$, $\delta_P = \omega_{24} - \omega_P$, $\delta_A = \omega_{34} - \omega_A$, $\rho_{ij} = |i\rangle\langle j|$, $\Omega_i$ is the Rabi frequency of the laser beam $\omega_i$ (i=1,2,3), and $\gamma_{ij}$ and $\rho_{ij}$ are, respectively, a decay rate and a density matrix element for the transition between states |i> and |j>. By adding an additional laser beam $\omega_A$ and a ground state |3> into the Λ−type model of EIT composed of states |1>, |2>, and |4>, we obtain the degenerate dark resonance. Theoretical studies in a tripod model have been intensively performed [17-21]. For a special case of no detuning ($\delta_A = \delta_C = 0$), the degenerate dark state (|d1> and |d2>) at line center (see also the first row of Fig. 3) is obtained analytically as follows:

$$|d1\rangle = c_1\{\Omega_A|1\rangle - \Omega_C|3\rangle\},$$
$$|d2\rangle = c_2\{\Omega_P\Omega_C|1\rangle + \Omega_P\Omega_A|3\rangle - (\Omega_C^2 + \Omega_A^2)|2\rangle\}$$
(2)

It is clear that the dark state |d2> becomes equivalent to the EIT mode or single dark state (|d>) if the control light $\Omega_A$ is removed: $|d\rangle = c(\Omega_P|1\rangle - \Omega_C|2\rangle)$. On the other hand, at both sidebands, the interaction Hamiltonian results in bright modes, |b$_\pm$>, showing absorption:

$$|b_\pm\rangle = \frac{1}{\Omega}(\Omega_C|1\rangle + \Omega_P|2\rangle + \Omega_A|3\rangle \pm \Omega|4\rangle),$$
(3)

where $\Omega = \sqrt{\Omega_C^2 + \Omega_A^2 + \Omega_P^2}$. The energy separation between two bright modes (i.e., absorption peaks) is determined by the generalized Rabi frequency $\Omega$.

Figures 3 show numerical simulations for both the metamaterial and tripod system of Fig. 2. Figure 3(a) (the left column) presents simulation results for the metamaterial, and Fig. 3(b) (the right column) presents results for the corresponding tripod system by using Eq. (1). As seen in the first row, the metamaterial with equal length ($L_1=L_2=L$) of symmetric double parallel metal stripes in Fig. 2 satisfies a degenerate tripod atomic system, which proves a degenerate engenvalue of Eqs. (2) and (3). The red dotted curve in the first row of Fig. 3 is for EIT in Fig. 1 (without the control field $\Omega_A$) as a reference. As expected, the peak to peak separation determined by $\Omega$ is wider than that by EIT by $\sim \sqrt{2}$ ($\Omega \sim \sqrt{2}\Omega_C$.) The general solution of Eq. (3) with a symmetric detuning ($\delta_C=-\delta_A=\delta$) can also be obtained from numerical simulations by introducing asymptotic relationship with a detuning $\delta$:

$$e_\pm = \pm \frac{\Omega}{\sqrt{2}} \sqrt{1+2\left(\frac{\delta}{\Omega}\right)^2}, \qquad (4)$$

where the peak to peak energy separation must be greater than that obtained from Eq. (3): See Figs. 3 and 4.

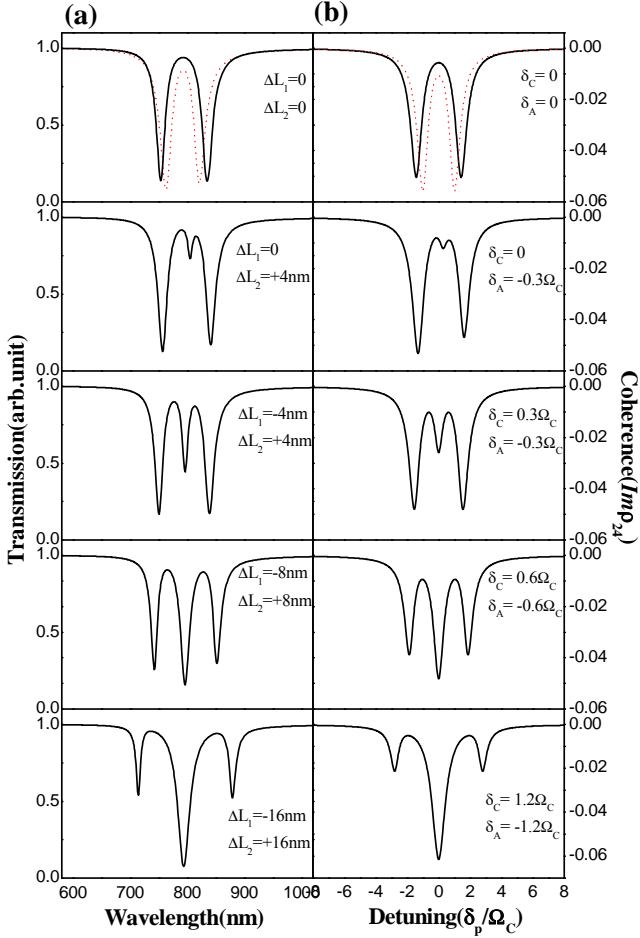

FIG. 3 (color online). (a) Numerical analysis for the plasmon-induced photonic switching in Fig. 2. The red curve in the first row is for Fig. 1(a) as a reference.

We now discuss double dark resonance, which gives an absorption enhancement otherwise dark state of Eq. (2) when a detuning $\delta$ is introduced. The length of the double parallel metal stripes $L_1$ and $L_2$ changes. As indicated above, the length change induces a two-photon detuning ($\delta_C$ or $\delta_A$) within the coupling Rabi frequency. For the rest of Fig. 3, all parameters of Fig. 1(a) remain the same unless otherwise indicated. For the second row of Fig. 3, where $L_1=L$ and $L_2=L+\Delta L$ ($\Delta L = 4$ nm corresponding to $\delta_A = -0.3\Omega_C$ and $\delta_C = 0$), a weak absorption enhancement appears near line center. Due to the detuning effect of the control $\Omega_A$, the spectral position of the double dark resonance shifts slightly to the right. Here the dark states (EIT transparency) appear at both $\delta_P = 0$ and $\delta_P = \delta_A$. The eigenvalue of the double dark state (enhanced absorption) is determined by the relationship among $\delta_A$, $\Omega_A$, and $\Omega_C$. On the third row of Fig. 3, the length of the parallel double metal stripes on the left wing is modified to be decreased by $\Delta L$ ($L_1=L-\Delta L$). Because the detuning is still within the Rabi frequency ($\delta_A = -0.3\Omega_C$; $\delta_C = 0.3\Omega_C$; the corresponding Rabi frequency is analyzed through the simulations), the quantum coherence can still be sustained by inducing the double dark resonance resulting in absorption enhancement at line center (This will be discussed in Fig. 4.) Compared with the second row, the frequency shift of the probe on the third row is zero due to the balanced detuning. As the length of the double parallel metal stripes continues to change, the system behaves noncoherently indicating broadened absorption linewidth at line center (see the fifth row): More detailed will be discussed elsewhere. All corresponding figures on the right column for a tripod atomic model match quite well those in the left column.

In Fig. 4, we discuss theoretical analysis of the present metamaterial for plasmon-induced photonic switching using coherence swapping or double dark resonance in a tripod model. First of all, we discuss the tripod model of Fig. 2 to determine if the absorption enhancement at line center is really due to quantum coherence. In the inset of Fig. 4 we assume that the tripod model (A) can be decomposed into two independently detuned EIT models (B and C). Figures 4(a) and 4(b) are for metamaterial and atomic system, respectively. Then the sum of both detuned EIT spectra is compared with that of the tripod model. As a result, each probe spectrum for the detuned model shows the mirror image of the other across the line center (see the red dotted and blue dash-dot curves in Figs. 4(a) and 4(b)). Compared with the third row of Fig. 3, no absorption enhancement occurs in the sum spectrum at line center in Figs. 4(a) and 4(b). Therefore, the assumption must be wrong. Figure 4(c) shows probe absorption as a function of coupling/control detuning $\delta$ for a broader optical homogeneous width: $\gamma_O > \Omega_C$, $\Omega_A$ ($\Omega_C = \Omega_A=10$ kHz; $\gamma_O=30$ kHz). At both side bands across zero detuning, an asymmetric feature of the absorption (blue) and transparency (red) is shown along the detuning $\delta$ as discussed in Eq. (4). In the region of small detuning $\delta$ ($\delta \ll \Omega_C$; denoted by the dotted circle), the absorption width is narrower than the natural linewidth $\Gamma_O$, where $\Gamma_O = 6$ kHz. This result strongly supports that the absorption enhancement in a tripod system is due to quantum coherence. Figure 4(d) shows coherence variations for $\delta=0.2\Omega_C$. Curves (i) and (ii) are references in the EIT system showing that optical coherence $Im\rho_{24}$ (i) correlates with two-photon coherence $Re\rho_{12}$ (ii). Curves (iii) and (iv) in a tripod system represent that the preexising two-photon coherence $Re\rho_{12}$ becomes transferred into a newly generated two-photon coherence $Re\rho_{23}$ between states $|2>$ and $|3>$. Thus, the optical transparency (EIT) in a $\Lambda$−type system turns into absorption enhancement in a tripod system due to spin coherence swapping. The spectral width of the optical switching can be controlled by adjusting the detuning $\delta$ as shown in Fig. 4(c).

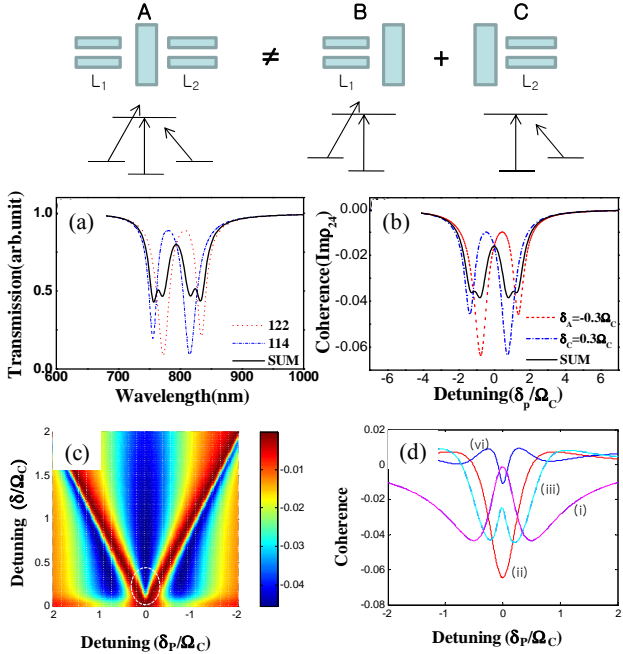

FIG. 4 (color online). Numerical analysis of plasmon-induced photonic switching based on coherence swapping. (a) For metameterial B [$L_1$ = 114 nm (blue dash dotted)] and C [$L_2$ = 122 nm (red dotted)]; averaged sum (black). (b) For corresponding EIT models of (a) with no correlation between them. (c) Coherence $Im\rho_{24}$ for a tripod system A as functions of probe detuning $\delta_P$ and $\delta$ ($\delta=\delta_C=-\delta_A$): $\Gamma_O=6$; $\gamma_O=30$; $\Omega_P=1$; $\Omega_C=\Omega_A=10$ kHz. (d) coherence versus probe detuning: For a $\Lambda$–type system of Fig. 1, (i) $Im\rho_{24}$ and (ii) $Re\rho_{12}$; For a tripod system of Fig. 2, (iii) $Re\rho_{12}+Re\rho_{13}$; and (iv) $Re\rho_{23}$; $\delta_C=-\delta_A=2$ kHz.

In conclusion, we have presented a plasmon-induced double-dark resonance for the switching phenomenon and analyzed it by using a tripod optical model. By modifying the plasmon model of EIT for a tripod system, we have successfully demonstrated plasmon-induced optical switching by using the coherence swapping phenomenon or double dark resonance observed in a tripod solid system based on EIT [17,19]. By applying the proposed metamaterial scheme of a tripod system to optical waveguides or surface plasmon polaritons, an active plasmon-induced photonic processing can be implemented. Furthermore, the present plasmonic mode control may benefit applications of a giant phase shift without signal loss even at weak field limit [21].

We acknowledge that this work was supported by the CRI program (Center for Photon Information Processing) of Korean Ministry of Education, Science and Technology via KOSEF. We thank S. Zhang and X. Zhang of UC Berkeley, and Y. H. Lu and Y. P. Lee of Hanyang University for helpful discussions.